\documentstyle[prc,twocolumn,aps,epsf]{revtex}
\begin{document}
\twocolumn[\hsize\textwidth\columnwidth\hsize\csname
@twocolumnfalse\endcsname

\draft
\title{Low-lying continuum structures in $\bbox{^8}$B and 
$\bbox{^8}$Li in a microscopic model}
\author{Attila Cs\'ot\'o}
\address{Department of Atomic Physics, E\"otv\"os University,
P\'azm\'any P\'eter s\'et\'any 1/A, H--1117 Budapest, Hungary}
\date{August 12, 1999}

\maketitle

\begin{abstract}
We search for low-lying resonances in the $^8$B and $^8$Li nuclei 
using a microscopic cluster model and a variational scattering 
method, which is analytically continued to complex energies. 
After fine-tuning the nucleon-nucleon interaction to get the 
known $1^+$ state of $^8$B at the right energy, we reproduce 
the known spectra of the studied nuclei. In addition, our model 
predicts a $1^+$ state at 1.3 MeV in $^8$B, relative to the 
$^7{\rm Be}+p$ threshold, whose corresponding pair is situated 
right at the $^7{\rm Li}+n$ threshold in $^8$Li. Lacking any 
experimental evidence for the existence of such states, it is 
presently uncertain whether these structures really exist or 
they are spurious resonances in our model. We demonstrate that 
the predicted state in $^8$B, if it exists, would have important 
consequences for the understanding of the astrophysically 
important $^7{\rm Be}(p,\gamma){^8{\rm B}}$ reaction.
\end{abstract}
\ \\
]

\narrowtext

\section{Introduction}

The $^7{\rm Be}(p,\gamma){^8{\rm B}}$ and 
$^7{\rm Li}(n,\gamma){^8{\rm Li}}$ reactions play important
roles in nuclear astrophysics. The former process produces
$^8$B in the sun, which is the dominant source of the
high-energy solar neutrinos (via the beta decay to
$^8$Be) \cite{Bahcall}, while the latter one is a key reaction
in inhomogeneous big bang models \cite{Rolfs}. In stellar
environments only the very low-energy continuum structures of
the participating nuclei are probed. However, terrestrial
experiments often cannot study these low-energy processes 
directly, and have to extrapolate the measured data (e.g.,
cross sections) from higher energies down to the
astrophysically interesting region. Obviously, the
extrapolation procedure can be strongly influenced by higher
lying structures present in the continuum. For example, the
known $1^+$ and $3^+$ resonances in $^8$B cause bumps in the
$^7{\rm Be}(p,\gamma){^8{\rm B}}$ cross section, which have to
be treated properly if one wants to interpret the results of
both the radiative capture \cite{expcap} and Coulomb
dissociation \cite{expdis} measurements correctly. 

The aim of the present work is to explore the continuum
structure of $^8$B and $^8$Li using a microscopic model that
has been previously applied to describe the ground states of 
these nuclei \cite{B8gs} and in the calculation of the
nonresonant $^7{\rm Be}(p,\gamma){^8{\rm B}}$ cross section
\cite{B8,B8offsh}. 

According to the Ref.\ \cite{Ajzenberg} compilation, there is
a resonance in $^8$B at $E_r=0.63$ MeV (in this paper the
energies are given in the center-of-mass frame, relative to
the $^7{\rm Be}+p$ or $^7{\rm Li}+n$ threshold) and a $3^+$
state at 2.18 MeV. Many calculations predict the first state
to be a $1^+$ resonance, see, e.g., Ref.\ \cite{calc}. In $^8$Li 
there is a bound $1^+$ state, a $3^+$ resonance at 0.22 MeV, 
and a broad $1^+$ state at 1.18 MeV. The Ref.\ \cite{Ajzenberg} 
compilation does not give any further low-lying states in 
$^8$Li and $^8$B. However, there are indications for the existence 
of a $2^-$ state in $^8$Li at 1.18 MeV \cite{Knox}, for a $1^+$ 
state of $^8$B at 2.7 MeV, and for a $1^-$ or $2^-$ level in 
$^8$B at 2.9 MeV \cite{Goldberg}. 

As far as theoretical models are concerned, the potential
model of Ref.\ \cite{Kim} seemed to indicate an additional M1
($1^+$ or $2^+$) state in $^8$B at about 1.4 MeV. However,
that calculation was criticized \cite{Barkercom} by pointing
out that certain model assumptions in \cite{Kim}, like the
$^7{\rm Be\,(g.s.)}+p$ nature of the M1 states, were rather
unphysical. Signs of new states have also surfaced in
microscopic models. For example, in Ref.\ \cite{B8offsh} we
noted that our microscopic $^4{\rm He}+{^3{\rm He}}+p$ cluster
model also seemed to indicate the presence of a $1^+$
resonance at around 1.5 MeV.

In order to be able to make a well-founded prediction on the
possible existence of new continuum states in $^8$B (and 
$^8$Li) in any model, one should satisfy two important 
criteria. First, the properties of the known $1^+$ state at
0.63 MeV must be well reproduced by the model. In certain
approaches, like in the potential model, this is a tough
problem. Secondly, the recognition of any possible new
resonance should be fairly reliable, even for broad states.
Below we present a model which satisfies both conditions and
predicts new states in $^8$B and $^8$Li.
 
\section{Model}

We use the same model as in Refs.\ \cite{B8gs,B8,B8offsh}. The
eight-body, three-cluster ($^4{\rm He}+{^3{\rm He}}+p$) wave
function of $^8$B is chosen as 
\begin{equation}
\Psi=\sum_{I_7,I,l_2}\sum_{i=1}^{N_7}
{\cal A}\Big \{\big [ \big [\Phi^p_{s_2}\Phi^{^7{\rm
Be},i}_{I_7} \big ]_I
\chi_{l_2}^i(\bbox{\rho }_2) \big ]_{JM} \Big \},
\label{wf}
\end{equation}
where ${\cal A}$ is the intercluster antisymmetrizer,
$\bbox{\rho}_2$ and $l_2$ are the relative coordinate and
relative angular momentum between $^7$Be and $p$,
respectively, $s_2$ and $I_7$ are the spin of the proton and
$^7$Be, respectively, $I$ is the channel spin, and [...]
denotes angular momentum coupling. While $\Phi^p$ is a proton
spin-isospin eigenstate, the antisymmetrized ground state
($i=1$) and continuum excited distortion states ($i>1$) of
$^7$Be are represented by the wave functions
\begin{equation}
\Phi^{^7{\rm Be},i}_{I_7}=\sum_{j=1}^{N_7}c_{ij}\sum_{l_1}
{\cal A}\Big \{\big [ \big [\Phi^\alpha\Phi^h \big
]_{s_1}\Gamma_{l_1}^j(\bbox{\rho }_1) \big ]_{I_7M_7}
\Big \}.
\label{wf1}
\end{equation}
Here ${\cal A}$ is the intercluster antisymmetrizer between
$\alpha$ and $h$, $\Phi^\alpha$ and $\Phi^h$ are
translationally invariant harmonic oscillator shell model
states ($\alpha={^4{\rm He}}$, $h={^3{\rm He}}$), 
$\bbox{\rho}_1$ is the relative coordinate between $\alpha$ 
and $h$, $l_1$ is the $\alpha -h$ relative angular momentum, $s_1$ 
is the spin of h, and
$\Gamma_{l_1}^j(\bbox{\rho_1})$ is a Gaussian function with a
width of $\gamma_j$. The $c_{ij}$ parameters are determined
from a variational principle for the $^7$Be energy. A similar
wave function is used for $^8$Li within the $^4{\rm 
He}+{^3{\rm H}}+n$ cluster model space. Using (\ref{wf}) in 
the 8-nucleon Scr\"odinger equation, we get an equation for 
the unknown relative motions $\chi$. 

In order to ensure that all existing resonances are recognized
in the continuum, we search for the poles of the $^7{\rm
Be}+p$ ($^7{\rm Li}+n$) scattering matrices. These matrices
are generated with the help of the Kohn-Hulth\'en variational 
method for scattering processes \cite{Kamimura}. In that
method the relative motions in (\ref{wf}) are expanded in
terms of square-integrable functions (Gaussians in our case)
matched with the correct scattering boundary condition.
The resulting variational scattering matrices are analytically
continued to the Riemann surface of complex energies, using the
methods of Ref.\ \cite{he5}. That is, we generate solutions to
the scattering problem with such asymptotic behavior that
corresponds to complex energies,
\begin{equation}
\chi(\varepsilon,\rho)
\rightarrow H^-(k\rho)-\widetilde S(\varepsilon) H^+(k\rho).
\end{equation}
Here $\varepsilon$ and $k$ are the {\it complex} energies
and wave numbers of the relative motions, and $H^-$ and
$H^+$ are the incoming and outgoing Coulomb functions,
respectively. These solutions have no physical meaning except
when $\widetilde S$ is singular, where $\widetilde S$ coincides with
the physical $S$ matrix. We perform this analytic continuation
procedure and search for the poles of $S$. The parameters of
the complex-energy poles are related to the resonance
parameters ($E_r$ position and $\Gamma$ width) through the 
\begin{equation}
\varepsilon=E_{\rm r}-i\Gamma/2
\end{equation}
equation. We mention that this way of determining resonance
parameters is the most closely related to scattering theory 
\cite{Newton}. The results obtained can differ from those
coming from other methods, especially for broad states
\cite{he5}.

We note that the $^7{\rm Be}+p$ scattering picture is valid
only below the $^4{\rm He}+{^3{\rm He}}+p$ three-body breakup
threshold, which is at 1.587 MeV experimentally (2.467 MeV in
the case of $^7$Li and $^4{\rm He}+{^3{\rm H}}+n$). Above this
energy three-body scattering asymptotics should be used,
which is beyond our capacity. Although the continuum-excited
distortion states in Eq.\ (\ref{wf}) can mimic three-body
break-up at some approximate level, any states predicted above
these energies in our model should be handled with care.

Our approach contains a large and physically motivated part of
the 8-body Hilbert space. The next step is to find an effective
nucleon-nucleon (N-N) interaction that is the most suitable 
for the description of the problem at hand. As we mentioned,
the precise reproduction of the known $1^+$ state in $^8$B
(and in $^8$Li) is a first priority. In most of our previous
calculations for $^8$B \cite{B8,B8offsh} we used the Minnesota
(MN) force. However, it turned out that this interaction gave
an incorrect channel spin ratio for the $1^+_1$ state of
$^8$Be \cite{be8}, which is a member of a $T=1$ triplet
together with the $1^+$ states of $^8$B and $^8$Li. This flaw
is most probably caused by the fact that the reproduction of 
this state at the right energy in $^8$Be requires an exchange
mixture parameter value ($u$) of the MN force which is
incompatible with the experimental data in the singlet-odd N-N
channel ($u>1$). To avoid this problem in the present work,
here we apply the modified Hasegawa-Nagata interaction
\cite{MHN}, which is also widely used in cluster model 
calculations.

We assume a pure Wigner form for the spin-orbit force and
fine-tune the Majorana component (while adjusting the Wigner
component to keep $W+M={\rm const.}$) of the medium range part
of the central interaction, in order to get the $1^+$ state of
$^8$B at 632 keV, its experimental value taken from Filippone
{\it et al.} \cite{expcap} (note that Ref.\ \cite{Ajzenberg}
gives 637 keV). The required modification in the central force
is about 5\%. The strength of the short-range spin-orbit
interaction is fixed by requiring that the experimental
spin-orbit splitting between the $3/2^-$ and $1/2^-$ states of
$^7$Be be reproduced correctly. We use the same interaction
for both $^8$B and $^8$Li. Our prime target is $^8$B here, a 
more precise reproduction of the $^7$Li states and the
fine-tuning of the N-N force for the $1^+$ resonance of $^8$Li
is not pursued.
 
\section{Results}

We show the spectrum of $^8$B and $^8$Li coming from our
calculation in Table \ref{tab1}, together with the
experimental numbers. In these calculations we used $l_1=1$ 
inside the 7-nucleon subsystems in (\ref{wf1}) in all cases, 
$l_2=1$ in the case of the $1^+$ and $2^+$ states, and 
$l_2=1$ and 3 in the case of the $3^+$ state. The total 
spin $S$ took all possible values. We emphasize
again that our model is not a fully adequate description for
those states that lie above the three-body thresholds. These
thresholds are situated at 1.34 MeV and 2.13 MeV, respectively
in our model, compared to the experimental 1.59 MeV and 2.47
MeV. One can see in Table \ref{tab1}, that the properties of
the $1^+_1$ state in $^8$B are nicely reproduced. This is a
significant achievement as other models, most notably the
$^7{\rm Be}+p$ potential model, tend to strongly overestimate
the width of this state. We also note that the N-N interaction
which is fine-tuned to get the $1^+$ state at the right
position, slightly overbinds the $2^+$ ground state. This can
be explained by the fact that the $({^4{\rm He}},{^3{\rm
He}})p$ model space is closer to the true wave function of 
the $2^+$ ground state than to that of the $1^+$ resonance. It
means that while describing the $1^+$ state, the N-N force has
to be made stronger in order to compensate for those missing
parts of the Hilbert space that cannot be represented by the 
$({^4{\rm He}},{^3{\rm He}})p$ wave function. However, this
modification of the force leads to the slight overbinding in 
the $2^+$ state. 

The most surprising result in Table \ref{tab1} is the
prediction of an additional $1^+$ state at 1.28 MeV. Note that
although this resonance is situated rather close to our 
$^4{\rm He}+{^3{\rm He}}+p$ three-body threshold, it is not a
threshold effect or a similar artifact. By making the N-N
force slightly stronger, both the energy of the $1^+_2$ state
and that of $^7$Be is lowered (while the position of the
three-body threshold is unchanged), which makes this state
more bound relative to the three-body threshold. The
corresponding $1^+_2$ state in $^8$Li is very close to the
$^7{\rm Li}+n$ two-body threshold, therefore its parameters
can depend on the details of the model rather strongly.

One can see in Table \ref{tab1} that the known $3^+$ states
appear in our model, although somewhat shifted to higher
energies, compared with the experimental situation. In
addition, we have a third $1^+$ state, although it lies above
the three-body threshold in both $^8$B and $^8$Li. Whether
this state would appear in a realistic model, which contains 
the correct three-body asymptotics, remains a question. 

Turning our attention back to the $1^+_2$ states, we note that
no experimental evidence has been found so far that would
support the existence of such structures \cite{Ajzenberg}. It
seems rather unlikely, e.g., that the existence of such a
state in $^8$Li, right at the $^7{\rm Li}+n$ threshold, could
be reconciled with the cross section measurements of Ref.\
\cite{Heil}. Also, no indications of such states were found 
in $^6{\rm Li}(t,p){^8{\rm Li}}$ and 
$^9{\rm Be}(d,{^3{\rm He}}){^8{\rm Li}}$ experiments 
\cite{Ajzenberg}. Nevertheless we feel obliged to report on the
existence of these states in our model. Although rather
unlikely, these states can be spurious solutions in our model.
Certain variational approaches are known to produce such
artifacts \cite{spur}. Spurious states can also appear if
incorrect boundary conditions are used. This would happen here
also in the case of $^8$Li, if we searched for $2^+$ states in
a three-body bound state calculation. We would find two such
three-body bound states, one of them lying above the $^7{\rm
Li}+n$ threshold. Using a wave function which has the correct
asymptotic behavior (scattering state in $^7{\rm Li}+n$) would
result in the disappearance of the second ``state''. For
further details, see \cite{comm}. The situation is different
in the present work. We do not know any conceivable reason why
spurious states might appear in our description. 
Another possibility is that the predicted states are situated 
at significantly higher energies in reality, but for some 
reason they appear at low energies in our model. In the 
following we discuss some characteristic features of the newly
found resonances, which might help to find them experimentally
if they exist, or else find a way to understand them and get
rid of them in our model if they turned out to be spurious
states.

In Table \ref{tab2} we give the probabilities of the various
angular momentum channels which build up the wave functions of
the $1^+$ states of $^8$B and $^8$Li in several different 
forms. In these calculations a three-body bound state
approximation was used for the $1^+$ resonances, where the
relative motions were expanded in terms of square-integrable
basis functions. The resulting probabilities characterize the
inner parts of the resonance wave functions rather well. 

First the weights of the orthogonal channels are given in
Table \ref{tab2} in the $(L,S)$ and $(I,l_2)$ representations, 
respectively. Here $[(l_1,l_2)L,(s_1,s_2)S]J$ and 
$[[(l_1,s_1)I_7,s_2]I,l_2]J$ coupling schemes are assumed,
respectively. The third way of expressing the importance of
the various {\it nonorthogonal} channels is to calculate their
amounts of clustering \cite{amcl}. This quantity gives the
probability that the full wave function of a state lies
completely in the given subspace (thus the sum of the
probabilities is obviously not 1). The fourth way of
characterizing the $1^+$ states in Table \ref{tab2} is to
calculate the weights of the shell-model-like $\vert
I_7,j\rangle$ configurations, where $I_7=3/2$ or 1/2 is the
spin of $^7{\rm Be}/{^7{\rm Li}}$, while $j=3/2$ or 1/2 comes
from the coupling of the orbital momentum ($l_2$) and spin 
($s_2$) of $p/n$. We calculated this quantity in the 
following way. First we expressed the $\vert I_7,p_j\rangle$ 
state in terms of the $(L,S)$ components, and then evaluated 
its square by substituting the $(L,S)$ weights.

As one can see in Table \ref{tab2}, the $1^+_1$ state receives
its main contribution from the $I=2$ channel spin components,
while $1^+_2$ is dominated by the $I=1$ channel spin. In
addition, as shown by the amounts of clustering, the $1^+_1$
state receives large contributions from configurations which
contain a $^7$Be in its excited state ($I_7=1/2$), while in
$1^+_2$ the role of the excited $^7$Be core is negligible.

The newly found $1^+_2$ states, if they exist, would have 
important consequences for the radiative capture reactions 
$^7{\rm Be}(p,\gamma){^8{\rm B}}$ and $^7{\rm Li}(n,
\gamma){^8{\rm Li}}$. We do not discuss these consequences in
detail, just mention one interesting example. In order to be
able to reliably extrapolate the experimental   
$^7{\rm Be}(p,\gamma){^8{\rm B}}$ cross section down to
stellar energies, the precise knowledge of the nonresonant E1
cross section is necessary. The $1^+_2$ resonance in the
vicinity of 1.3 MeV, if it exists but is not recognized, would
change the average slope of the extracted E1 cross section.
In order to show this effect, we calculated the M1 cross
section using the above model. Due to technical
simplifications, the asymptotic behavior of the $2^+$ bound
state could only be correctly described up to $15-20$ fm. Note
that this is an acceptable approximation (in contrast to the
E1 cross section, where it would be totally unphysical) at
least in the resonance regions, where the scattering wave
functions are large at small radii. 

\begin{figure}
\centerline{\epsfxsize 12cm \epsfbox{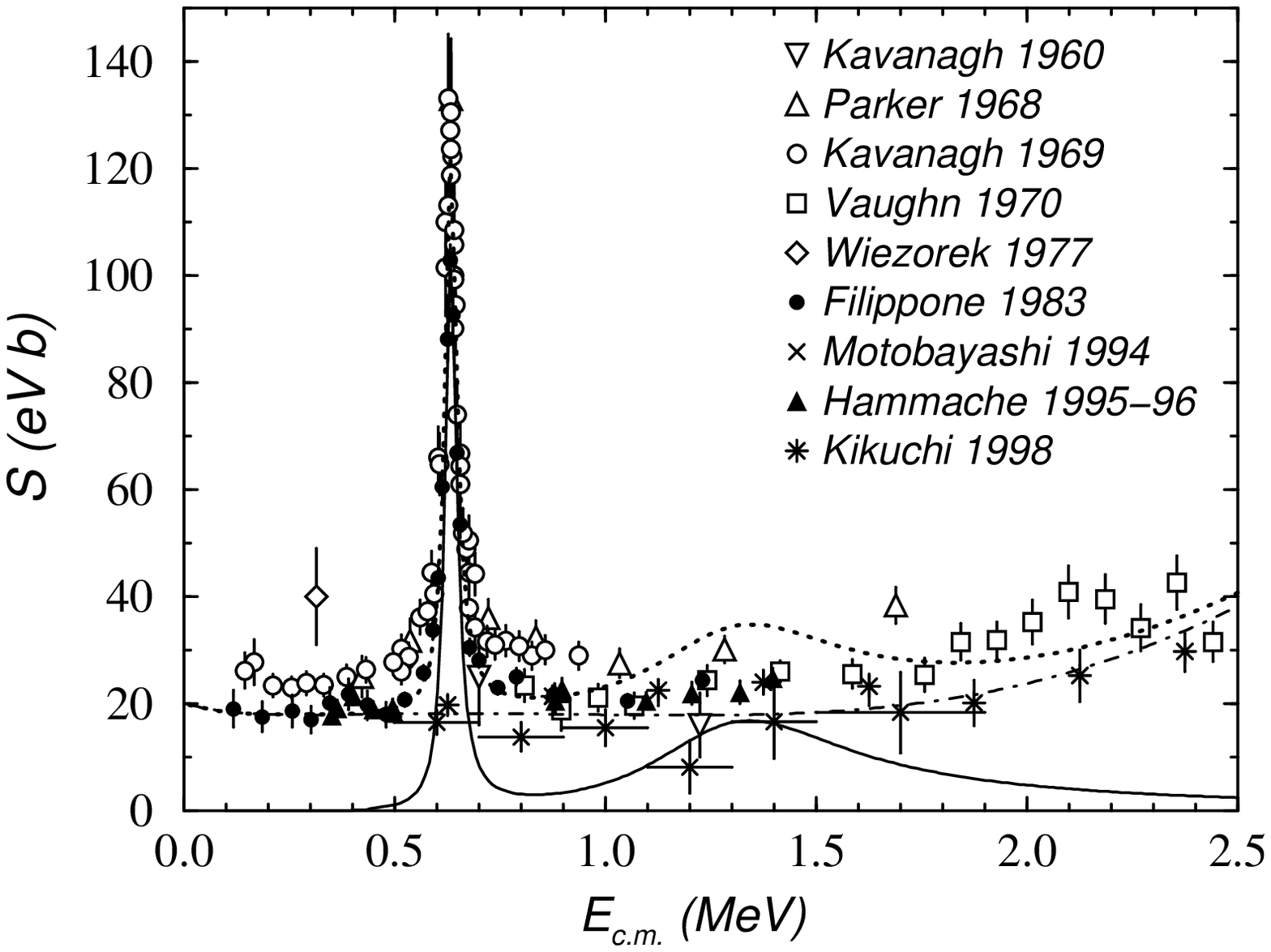}}
\caption{Astrophysical S factor for the $^7{\rm Be}(p,
\gamma){^8{\rm B}}$ reaction. The solid curve shows the M1
cross section coming from the $1^+$ initial states in the
present model. The dot-dashed curve is an E1 cross section
whose energy dependence is taken from \protect\cite{B8,B8offsh}
and its absolute normalization is fitted (by eye) to the
low-energy Filippone data. The dotted curve is the sum of the
two. Also shown are the experimental data 
\protect\cite{expcap,expdis} with $\sigma_{dp}=157$ mb 
normalization, where it applies.}
\label{fig1}
\end{figure}

The resulting M1 cross section, coming from the $1^+$ initial
scattering states (that is, no $3^+$ state is included), is
shown in Fig.\ \ref{fig1} (solid line). Also shown in Fig.\
\ref{fig1} is the E1+M1 cross section (dotted curve) which we
get by adding our M1 result to an E1 curve (dot-dashed line)
which has the energy dependence of Refs.\ \cite{B8,B8offsh}
and is fitted (by eye) to the low-energy Filippone data 
\cite{expcap}. One
can see that, although the $1^+_2$ bump predicted by our model
is much too strong, such a structure in this energy region is
not a priori ruled out by the data. A broader and smaller bump
might be consistent with the experiments. Therefore, if such a
state really exists, then it is weaker than our model
prediction in Table \ref{tab1}. We note that the biggest
contribution to the $1^+_2$ bump in Fig.\ \ref{fig1} comes
from the initial scattering states where the spin of $^7$Be is
$I_7=3/2$ and the channel spin is $I=1$. 

As one can observe, the average slope of the dotted curve in
Fig.\ \ref{fig1} is rather different from that of the E1 cross
section in the $0.8<E<1.5$ MeV region. This means that if the
$1^+_2$ resonance really existed in $^8$B, then it would make
the separation of the measured cross section into E1 and M1
components more difficult than currently believed.
 
\section{Conclusions}

In summary, we have searched for resonances in the $^8$B and
$^8$Li continua, using a microscopic three-cluster model. We
used a method, the analytic continuation of the scattering
matrix to complex energies, which is well suited to find any
resonances, even broad structures, if they exist. The
properties of the known $1^+$ states in $^8$B and $^8$Li are
well reproduced in our model. In addition, we have found new
$1^+$ resonances. We have shown that the state predicted at
1.3 MeV in $^8$B would have important consequences in the
understanding of the $^7{\rm Be}(p,\gamma){^8{\rm B}}$
reaction. 

As currently there is no experimental indication for the
existence of the predicted states, we cannot be sure whether these
structures are real or not. In order to be able to confirm or
refute the existence of the predicted states, more theoretical
and experimental work would be needed.

\acknowledgments

This work was supported by OTKA Grants F019701, F033044, 
and D32513, and by the Bolyai Fellowship of the Hungarian 
Academy of Sciences. We thank Fred Barker for some useful 
comments and for calling our attention to the Ref.\ 
\cite{Heil} work.

\newpage 

\narrowtext
\begin{table}
\caption{The low-energy spectrum of $^8$B and $^8$Li. 
The experimental data are taken from \protect\cite{Ajzenberg},
except for the first $1^+$ state of $^8$B, which is from
Filippone \protect\cite{expcap}. All numbers are in MeV.}
\begin{tabular}{ccr@{}lr@{}lr@{}lr@{}l}
 & & \multicolumn{4}{c}{Theory} & 
                           \multicolumn{4}{c}{Experiment}  \\
\cline{3-6}
\cline{7-10}
 & $J^\pi$ & \multicolumn{2}{c}{$E_r$} & \multicolumn{2}{c}{$\Gamma$} &
 \multicolumn{2}{c}{$E_r$} & \multicolumn{2}{c}{$\Gamma$} \\
\tableline
$^8$B & $2^+$ & $-$0.&215 \ \ \ & & & \ \ \ \ $-$0.&137 & & \\
      & $1^+$ & 0.&632 & 0.&034 \ & 0.&632$\pm$0.01 &
                                 \  0.&037$\pm$0.005 \\ 
      & $1^+$ & 1.&278 & 0.&564 &  &  &  &  \\
      & $3^+$ & 2.&98  & 0.&808 & 2.&183$\pm$0.03 &
                                          0.&35$\pm$0.04  \\
      & $1^+$ & 4.&33  & 1.&5 &  &  &  &  \\    
\tableline
$^8$Li& $2^+$ & $-$2.&021 & &    & $-$2.&033 & & \\
      & $1^+$ & $-$0.&975 & &    & $-$1.&052$\pm$0.001 & & \\ 
      & $1^+$ & 0.&037 & 0.&006 &  &  &  &  \\
      & $3^+$ & 0.&937 & 0.&327 & 0.&222$\pm$0.003 &
                                       0.&033$\pm$0.006  \\
      & $1^+$ & 2.&29  & 1.&0 &  1.&18  & 
                            \multicolumn{2}{c}{$\sim\,$$1.0$} \\
\end{tabular}
\label{tab1}
\end{table}

\vskip 1.5cm

\mediumtext
\begin{table}
\caption{Characterization of the calculated $1^+$ states shown
in Table \protect\ref{tab1} by i) the weights of the 
orthogonal $(L,S)$ components, ii) the weights of the 
orthogonal $(I,l_2)$ components, iii) the amounts of 
clustering of the nonorthogonal $(I_7,I,l_2)$ components, and 
iv) the weights of the shell-model-like $(I_7,j)$ 
configurations in the $^8$B and $^8$Li wave function. For 
further details, see the text.}
\begin{tabular}{ccr@{}lr@{}lr@{}lr@{}lr@{}lr@{}l}
 & & \multicolumn{6}{c}{$^8$B} & \multicolumn{6}{c}{$^8$Li} \\
\cline{3-8}
\cline{9-14}
\multicolumn{2}{c}{Configuration} & 
                    \multicolumn{2}{c}{$1^+_1$} & \multicolumn{2}{c}{$1^+_2$} 
 & \multicolumn{2}{c}{$1^+_3$} & \multicolumn{2}{c}{$1^+_1$} & 
 \multicolumn{2}{c}{$1^+_2$} & \multicolumn{2}{c}{$1^+_3$} \\
\tableline
$(L,S)$ & $(0,1)$ & 0.&010 & 0.&001 & 0.&791 & 0.&001 & 
   0.&0002 & 0.&737 \\
 & $(1,0)$ & 0.&188 & 0.&728 & 0.&049 & 0.&211 & 
   0.&726 & 0.&077 \\
 & $(1,1)$ & 0.&786 & 0.&207 & 0.&154 & 0.&759 & 
   0.&237 & 0.&179 \\
 & $(2,1)$ & 0.&016 & 0.&064 & 0.&006 & 0.&029 & 
   0.&037 & 0.&007 \\
\tableline
$(I,l_2)$ & $(1,1)$ & 0.&237 & 0.&888 & 0.&184 & 0.&239 & 
   0.&870 & 0.&142 \\
 & $(2,1)$ & 0.&465 & 0.&107 & 0.&736 & 0.&417 & 
   0.&115 & 0.&849 \\
 & $(0,1)$ & 0.&298 & 0.&005 & 0.&080 & 0.&344 & 
   0.&015 & 0.&009 \\
\tableline
$(I_7,I,l_2)$ & $(3/2,1,1)$ & 0.&021 & 0.&890 & 0.&113 & 
   0.&030 & 0.&887 & 0.&138 \\
 & $(3/2,2,1)$ & 0.&621 & 0.&119 & 0.&816 & 
   0.&611 & 0.&195 & 0.&853 \\
 & $(1/2,1,1)$ & 0.&438 & 0.&048 & 0.&050 & 
   0.&474 & 0.&070 & 0.&004 \\
 & $(1/2,0,1)$ & 0.&472 & 0.&017 & 0.&054 & 
   0.&538 & 0.&051 & 0.&008 \\
\tableline
$(I_7,j)$ & $(3/2,3/2)$ & 0.&108 & 0.&409 & 0.&321 & 
   0.&123 & 0.&406 & 0.&316 \\
 & $(3/2,1/2)$ & 0.&418 & 0.&190 & 0.&317 & 
   0.&408 & 0.&202 & 0.&317 \\
 & $(1/2,3/2)$ & 0.&418 & 0.&190 & 0.&317 & 
   0.&408 & 0.&202 & 0.&317 \\
 & $(1/2,1/2)$ & 0.&049 & 0.&206 & 0.&044 & 
   0.&068 & 0.&191 & 0.&049 \\
\end{tabular}
\label{tab2}
\end{table}

\onecolumn
\newpage 

\mediumtext


\begin{references}
\bibitem{Bahcall} J.~N. Bahcall, {\it Neutrino
astrophysics} (Cambridge University Press, Cambridge,
1989).
\bibitem{Rolfs} C.~E. Rolfs and W.~S. Rodney, {\it
Cauldrons in the Cosmos} (The University of Chicago Press,
Chicago, 1988).
\bibitem{expcap} R.~W. Kavanagh, Nucl. Phys. {\bf 15},
411 (1960); P.~D. Parker, Phys. Rev. {\bf 150}, 851
(1966); Astrophys. J. {\bf 153}, L85 (1968); R.~W. Kavanagh, 
T.~A. Tombrello, J.~M. Mosher, and D.~R. Goosman, Bull. 
Am. Phys. Soc. {\bf 14}, 1209 (1969); R.~W. Kavanagh, in: 
{\it Cosmology, Fusion, and Other Matters}, Ed.\ F. Reines 
(Colorado Associated University Press, Boulder, 1972) p.\ 169;
F. J. Vaughn, R. A. Chalmers, D. Kohler, and L. F. Chase, 
Phys. Rev. C {\bf 2}, 1657 (1970); C. Wiezorek, H. 
Kr\"awinkel, R. Santo, and L. Wallek, Z. Phys. {\bf A282}, 121 
(1977); B. W. Filippone, A. J. Elwyn, C. N. Davids, and D. D. 
Koetke, Phys. Rev. Lett. {\bf 50}, 412 (1983); Phys. Rev. 
C {\bf 28}, 2222 (1983); F. Hammache {\it et al.}, Phys. Rev.
Lett. {\bf 80}, 928 (1998).
\bibitem{expdis} Motobayashi {\it et al.}, Phys. Rev.
Lett. {\bf 73}, 2680 (1994); N. Iwasa {\it et al.}, Journ. 
Phys. Soc. Japan {\bf 65}, 1256 (1996); T. Kikuchi {\it et 
al.}, Eur. J. Phys. {\bf 3}, 213 (1998).
\bibitem{B8gs} A. Cs\'ot\'o, Phys. Lett. B {\bf 315}, 24 (1993). 
\bibitem{B8} A. Cs\'ot\'o, K. Langanke, S.~E. Koonin, and 
T.~D. Shoppa, Phys. Rev. C {\bf 52}, 1130 (1995); A. Cs\'ot\'o 
and K. Langanke, Nucl. Phys. {\bf A636}, 240 (1998).
\bibitem{B8offsh} A. Cs\'ot\'o, Phys. Lett. B {\bf 394}, 247 
(1997).
\bibitem{Ajzenberg} F. Ajzenberg-Selove, Nucl. Phys. 
{\bf A490}, 1 (1988).
\bibitem{calc} P. Descouvemont and D. Baye, Nucl. Phys. 
{\bf  A487}, 420 (1988); Nucl. Phys. {\bf A567}, 341 (1994); 
Phys. Rev. C {\bf 60}, 015803 (1999). 
\bibitem{Knox} H.~D. Knox, D.~A. Resler, and R.~O. Lane, 
Nucl. Phys. {\bf A466}, 245 (1987). 
\bibitem{Goldberg} V.~Z. Gol'dberg, G.~V. Rogachev, M.~S. Golovkov, 
V.~I. Dukhanov, I.~N. Serikov, and V.~A. Timofeev, JETP Lett. 
{\bf 67}, 1013 (1998). 
\bibitem{Kim} K.~H. Kim, M.~H. Park, and B.~T. Kim, Phys.
Rev. C {\bf 35}, 363 (1987).
\bibitem{Barkercom} F. Barker, Phys. Rev. C {\bf 37}, 2920
(1988).
\bibitem{Kamimura} M. Kamimura, Prog. Theor. Phys. Suppl.
{\bf 62}, 236 (1977).
\bibitem{he5} A. Cs\'ot\'o, R.~G. Lovas, and A.~T. Kruppa,
Phys. Rev. Lett. {\bf 70}, 1389 (1993); A. Cs\'ot\'o and 
G.~M. Hale, Phys. Rev. C {\bf 55}, 536 (1997); Nucl. Phys. 
{\bf A631}, 783c (1998).
\bibitem{Newton} R.~G. Newton, {\it Scattering Theory of
Waves and Particles} (Springer, New York, 1982).
\bibitem{be8} A. Cs\'ot\'o and S. Karataglidis, Nucl. Phys. 
{\bf A607}, 62 (1996).
\bibitem{MHN} F. Tanabe, A. Tohsaki, and R. Tamagaki, Prog.
Theor. Phys. {\bf 53}, 677 (1975).
\bibitem{Heil} M. Heil, F. K\"appeler, M. Wiescher, and A.
Mengoni, Astrophys. J. {\bf 507}, 997 (1998).
\bibitem{spur} A. Cs\'ot\'o, B. Gyarmati, and A.~T. Kruppa, 
Few-body Syst. {\bf 11}, 149 (1991); E.~W. Schmid and 
J. Schwager, Nucl. Phys. {\bf A180}, 434 (1972); K. Lad\'anyi 
and T. Szondy, Nuovo Cim. {\bf 5B}, 70 (1971); R.~K. Nesbet, 
Phys. Rev. {\bf 175}, 134 (1968). 
\bibitem{comm} A. Cs\'ot\'o and R.~G. Lovas, Phys. Rev. C {\bf 
53}, 1444 (1996).
\bibitem{amcl} R. Beck, F. Dickmann, and R.~G. Lovas, Ann.
Phys. {\bf 173}, 1 (1987); A. Cs\'ot\'o and R.~G. Lovas, Phys. Rev.
C {\bf 46}, 576 (1992).
\end{references}
\end{document}